\documentclass[a4paper,12pt]{article}
\usepackage{fullpage}
\usepackage{graphicx}
\usepackage{amsmath,amsfonts}
\usepackage{caption}
\usepackage{subcaption}
\usepackage{float}
\usepackage{multicol}
\usepackage{subcaption}
\usepackage{setspace}
\usepackage[pagewise]{lineno}

\begin{document}

\title{Development of Conditional Random Field Insert for UNet-based Zonal Prostate Segmentation on T2-Weighted MRI }
\author{Peng Cao $^1$, \and
        Susan M. Noworolski $^1$, \and 
        Olga Starobinets $^1$, \and
        Natalie Korn $^1$, \and
        Sage P. Kramer $^1$, \and
        Antonio C. Westphalen $^1$,\and
        Andrew P. Leynes $^1$, \and
        Valentina Pedoia $^1$, \and
        Peder Larson $^1$ }
\date{ $^1$ Department of Radiology and Biomedical Imaging, University of California, San Francisco\\}


\maketitle

\noindent
$^*$ Corresponding to:\\
Peng Cao\\
Department of Radiology and Biomedical Imaging, University of California at San Francisco, San Francisco, CA, USA \\
Address: 1700 4th Street, San Francisco CA 94158\\
Email: caoxjtu@gmail.com\\

\noindent
\textbf{Short Running Title:}  Deep Learning for MRI Zonal Prostate Segmentation 
\\
\\

\newpage
 
\section*{Abstract}
\textbf{Purpose:} A conventional 2D UNet convolutional neural network (CNN) architecture may result in ill-defined boundaries in segmentation output. Several studies imposed stronger constraints on each level of UNet to improve the performance of 2D UNet, such as SegNet. In this study, we investigated 2D SegNet and a proposed conditional random field insert (CRFI) for zonal prostate segmentation from clinical T2-weighted MRI data. \\
\textbf{Methods:} We introduced a new methodology  that combines SegNet and CRFI to improve the accuracy and robustness of the segmentation. CRFI has feedback connections that encourage the data consistency at multiple levels of the feature  pyramid. On the encoder side of the SegNet, the CRFI combines the input feature maps and convolution block output based on their spatial local similarity, like a trainable bilateral filter. For all networks, 725 2D images (i.e., 29 MRI cases) were used in training; while, 174 2D images (i.e., 6 cases) were used in testing. \\
\textbf{Results:} The SegNet with CRFI achieved the relatively high Dice coefficients (0.76, 0.84, and 0.89) for the peripheral zone, central zone, and whole gland, respectively. Compared with UNet, the SegNet+CRFIs segmentation has generally higher Dice score and showed the robustness in determining the boundaries of anatomical structures compared with the SegNet or UNet segmentation. The SegNet with a CRFI at the end showed the CRFI can correct the segmentation errors from SegNet output, generating smooth and consistent segmentation for the prostate.\\ 
\textbf{Conclusion:} UNet based deep neural networks demonstrated in this study can perform zonal prostate segmentation, achieving high Dice coefficients compared with those in the literature. The proposed CRFI method can reduce the fuzzy boundaries that affected the segmentation performance of baseline UNet and SegNet models.

\clearpage
\section*{Introduction}
MRI of the prostate helps to localize cancer and characterize disease severity \cite{Wefer2000,Haider2007,Graser2007}. The prostate is composed of three regions: the central zone (CZ), the transition zone (TZ), and the peripheral zone (PZ). Approximately 70\% of cancers arise in the PZ, and 30\% in the TZ, with very few in the CZ \cite{pi-rads}. 
A PIRADS v2 score is given based on the zonal location of a visible lesion \cite{Hassanzadeh2017, Steiger2016}. For example, in the PZ the predominant MRI sequence used to estimate aggressiveness is diffusion-weighted imaging (DWI), but T2-weighted imaging (T2w) has greater weight when lesions are seen in the TZ \cite{Hassanzadeh2017, Steiger2016}. Therefore, an accurate prostate zonal segmentation could serve as the first processing step in the future automatized clinical decision-making pipeline: 1) MRI; 2) zonal segmentation; 3) lesion detection; 4) PIRADS grading.\par
Automated prostate segmentation provides essential and efficient tools for accurate measurements, primarily because manual segmentation is impractical for analyzing large 2D/3D datasets. Prostate segmentation research has been focused on the extraction of the entire gland for many years and achieved a Dice coefficient/accuracy of 0.90 in state-of-the-art algorithms \cite{Litjens2014}. For zonal prostate segmentation, conventional methods, such as template-based and C-means clustering \cite{Chilali2016} and level-set \cite{Toth2014}, provided the best Dice coefficients of 0.60 and 0.70 for PZ and combined CZ/TZ, respectively. However, these conventional zonal segmentation methods typically involved complex pre-processing, such as bias field correction, limiting their clinical application.\par
Recently, deep learning and convolutional neural networks (CNNs) have been applied to a broad range of medical imaging segmentation tasks such as brain, heart, lung, and prostate \cite{Ker2018}. The most widely used deep neural network in medical imaging segmentation is a fully convolutional network (FCN) \cite{Long2015} and its variations, including 2D and 3D UNet \cite{Cicek2016,Ronneberger2015} and 3D VNet \cite{Milletari2016} architectures. These FCNs provide both large receptive field and multi-scale representation for the input image, generating edge-preserved segmentations. For prostate segmentation, both UNet and VNet were successfully applied to the whole prostate segmentation \cite{Zhu2017, Milletari2016}, and achieved a Dice coefficient of 0.87 to 0.89 that was close to the performance of state-of-the-art algorithms \cite{Litjens2014}.  Meanwhile, the ill-defined boundary of segmentation output from 2D UNet was reported in reference \cite{Zhu2017}, which is likely related to the inadequate supervision in each internal level of UNet, as proposed by Zhu et al. \cite{Zhu2017}. Several recent improvements for the UNet can be generally considered as imposing stronger constraints on each level of UNet, e.g., providing multi-level “supervision” or ”attention focusing” \cite{Zhu2017, Oktay2018a}, i.e., attention-UNet, or using a fixed/simplified de-convolution kernel \cite{Badrinarayanan2017}, i.e., SegNet. Meanwhile, the conventionally widely used segmentation method, conditional random field (CRF), can also be applied to the segmentation output from CNN or FCN, i.e., CNN/FCN+CRF, resolving the sharp edge from blurred segmentation output from CNN or FCN \cite{Krahenbuhl2012, Teichmann2018}. A recent study recast the iterative inference process for CRF as a recurrent neural network (RNN) and introduced an end-to-end trainable FCN+CRFasRNN approach \cite{Zheng2015a}. However, these CRF approaches were computationally expensive, therefore, can not be inserted into the current UNet structure. In order to solve the ill-defined boundary for the UNet segmentation with CRF, one should simplify the current CRF implementation and made it compatible with the CNN architecture. 
In this study, we investigated 2D UNet and CRF inserts for zonal prostate segmentation from T2-weighted MRI data. We also introduced a new methodology that combines SegNet and the CRF insert (CRFI) to solve the ill-defined boundary from UNet segmentation. 

\section*{Methods}
\subsection*{Objects and MRI acquisition}
This study was approved by the Institutional Review Board at this institution and was compliant with the Health Insurance Portability and Accountability Act. Data from thirty-five participants, yielding a total of 875 2D images (i.e., 35 3D volumes) were used in this study. All participants provided written, informed consent and had a biopsy-confirmed diagnosis of prostate cancer.  \par
All participants were imaged with an expandable balloon endorectal coil (MedRad, Bayer HealthCare LLC, Whippany, NJ, USA) combined with an external phased-array coil on a 3T MR scanner (GE Healthcare, Waukesha, WI, USA). A perfluorocarbon fluid (Galden; Solvay Plastics, West Deptford, NJ, USA) was used to inflate the balloon coil. Fast spin-echo (FSE) T2-weighted images were acquired in an oblique axial plane with FOV = 18 cm, slice thickness = 3 mm, matrix = $256 \times 256$, and TR/TE = 6000/96 ms. The other acquisition details can be found in \cite{Starobinets2017}.\par

\subsection*{UNet and SegNet implementations} As shown in Figure \ref{fig:fig1}a, the convolutional neural network was created based on a UNet \cite{Ronneberger2015} structure with three decomposition levels. On the encoder side, i.e., on the contracting path, three $8\times 8$ convolutions (zero-padded convolutions) were used in each CNN block, and a $2\times 2$ max-pooling operation was used for the downsampling. Each convolution layer has 32 feature channels with a rectified linear unit (ReLU) activation function applied after each one. On the decoder side, i.e., on the expansive path, a $2\times 2$ transposed convolution was used for up-sampling, followed by a concatenation operation that combines the feature channels at the same level from the encoder side and the up-sampled feature channels from the lower level. The final layer was a $1\times 1$ convolution that was used to map each 32-channel feature vector to the three-class (background, PZ, and TZ) output. In total the network has 19 convolutional layers. Full-size 2D images/labels with a matrix size of $256 \times 256$ were used as the inputs/outputs for the network. The SegNet implementation was based on the abovementioned UNet, using a fixed/simplified de-convolution kernel with the max-pooling index copied from the same level on the encoder side
\cite{Badrinarayanan2017}. \par
\subsection*{CRFI implementation}
The CRF frame was based on the method in previous studies those used recurrent neural network to perform the inference for CRF \cite{Zheng2015a,long2018}. Briefly, CRF combined both unary dependencies between output features, $Y=\{y_1,y_2,...,y_N\}$, where N the number of pixels, and the input features for convolution block, $I=\{x_1,x_2,...,x_N\}$, as well as the pairwise dependencies between pairs of input features to produce the conditional probability $P(y_i|I)$, where $y_i$ is output feature vector.  $y_i$ was the label assigned to the pixel i, which was drawn from a pre-defined set of labels $\mathcal{L} =\{l_1, l_2,..., l_L\}$.
\begin{equation}
	P(y_i|I)= \dfrac{1}{Z} exp[-\sum_i \psi_U(y_i)-\sum_{i<j}\psi_P(y_i,y_j)]
	\label{eq:1}
\end{equation}
where  $\psi_U(y_i)$ the negative of the unary energy that encouraged the pixel-wise inverse likelihood of the conditional probability, $\psi_P(y_i,y_j)$ pairwise energy that promoted assigning same output feature for similar $y_i$ and $y_j$ \cite{Zheng2015a}, and $Z$ the normalizing constant. We followed the same setting as was used in \cite{Zheng2015a}, the negative of the unary energy was from the output of the previous convolution block. The additional pairwise energy promoted the combination of adjacent and similar feature vectors for generating smooth and consistent output feature vectors. In this paper, we defined the vector similarity between $x_i$ and $x_j$ as:

\begin{equation}
	S(x_i,x_j) = exp[-d(x_i,x_j)^2]
	\label{eq:2}
\end{equation}
where $d(x_i,x_j)$ the geometric distance between two vectors. We used the Eq. \eqref{eq:2} as an activation function that produced the similarity, $S = exp(-d^2)$, approximating the spatially Gaussian kernel. We further used a convolution layer to compute the effective/average distance $d$ for adjacent pixels, i.e., 
\begin{equation}
	d=Conv(I)
	\label{eq:3}
\end{equation}
where $I$ the input feature maps. For a small convolution kernel, i.e., $5\times 5$ in this study, the average distance can be a reasonable approximation of the pairwise distance between pixels. This was based on the fact that the convolution network (without the activation) can perform a linear operation locally on input feature vectors, e.g., average the local distance of the adjacent pixels or feature vectors. As a convolution kernel, in Eq. \eqref{eq:3}, it can perform the average and subtraction for every pixel surrounding the center pixel, and one can train a $5\times 5 \times K$ to $ 1$ convolution kernel, where $K$ the number of input features, resulting in one average local distance map, $d$. Then the vector activation function, in Eq. \eqref{eq:2}, can convert the distance map to a weighting map, which approximated a truncated local Gaussian kernel. Such  Gaussian similarity kernel can be simply calculated by a trainable convolution layer plus a new activation function in Eq.\eqref{eq:2}. \par
Similar to the method used in \cite{Zheng2015a,long2018}, two input feature vectors, e.g., $x_i$ and $x_j$, if they were similar in Eq. \eqref{eq:2} should be labeled with the similar or the same output feature vector, using a specially designed pairwise potential in Eq. \eqref{eq:1}. In this study, we followed the implementation of CRFasRNN, and the pairwise potential was defined as a weighted sum of Gaussian similarity kernel in Eq. \eqref{eq:2} \cite{Zheng2015a,long2018}:

\begin{equation}
	\psi_P(y_i,y_j) = \mu(y_i,y_j) \sum_m w^{(m)} S_m(x_i,x_j)
	\label{eq:4}
\end{equation}

where $m = 1,..., M$, the index of Gaussian kernel $S_{m}(.,.)$ that applied on feature vectors \cite{Zheng2015a}. In \cite{Zheng2015a}, function $\mu(., .)$, was the label compatibility function that adjusted label mutual weightings according to their compatibility, i.e., for $y_i=l,y_j=l'$ and $ l ,l'\epsilon \mathcal{L}$, $\mu(l,l')$ learned the mutual weightings between different labels regardless of spatial locations of $y_i$ and $y_j$. Therefore, maximizing the conditional probability in Eq. \eqref{eq:1} resulted in the most probable output feature vectors (those were chosen from the label set $\mathcal{L}$) based on combinations of similar input feature vectors \cite{Zheng2015a}.  \par
The implementation of CRF was based on the mean-field iteration method as an RNN that was initially proposed in \cite{Zheng2015a}. The iteration minimized the KL-divergence of the targeting probability distribution $P$ and the estimated $Q$ \cite{Zheng2015a}, which is given as the product of independent marginal distributions, i.e., 

\begin{equation}
	Q = \prod_i Q_i \\
	Q_i(y_i) = \sum_{y_j} \mu(y_i,y_j) \sum_m w^{(m)} S_m(x_i,x_j) Q_i(y_i) 
	\label{eq:5}
\end{equation}

In that study, authors recast the mean-field iteration method into a few iterative steps those can be embedded into the deep neural network, and they were: message passing, pre-weighting, compatibility transform, unary addition, and normalization \cite{Zheng2015a}. It should be noted that all steps in the original CRFasRNN can be readily implemented as a convolution layer and softmax activation function, except the message passing step was not compatible with the typical convolution layer  \cite{Zheng2015a}. The mean-field iteration method can be written as the below steps  \cite{Zheng2015a} with the modification on the Gaussian kernel computation in Eq. \eqref{eq:4}:
\\

\noindent
{Algorithm 1, Mean-field iteration for CRF in \cite{Zheng2015a} with similarity weighting defined in Eq. \eqref{eq:2}. } \\
$ U(l) = Conv\_block(I(l))$ \\ 
$ Q_i(l) = \dfrac{1}{Z_i} exp(U_i(l)) {, for all i} $  \null\hfill Initialization \\
$ {While (not\_converged)} $\\ 
$ \tilde{Q_i}^{(m)}(l) \leftarrow \sum_{j\neq i} S_{m}(x_i,x_j) Q_i(l){, for all m} $ \null\hfill Message passing\\
$ \check{Q_i}(l) \leftarrow \sum_m w^{(m)} \tilde{Q_i}^{(m)}(l)$ \null\hfill Pre-weighting \\
$ P_i(l) \leftarrow \sum_{l'\epsilon \mathcal{L}} \mu(l,l') \check{Q_i}(l)$ \null\hfill Compatibility transform \\
$ \tilde{Q_i}(l) \leftarrow U_i(l)-P_i(l)$                        \null\hfill Add unary\\
$ Q_i \leftarrow \dfrac{1}{Z_i} exp(\tilde{Q_i}(l))$                      \null\hfill Softmax  \\
\label{alg:1}

\begin{figure}
\centering
\begin{tabular}{c}
\begin{subfigure}[b]{0.9\textwidth}
\includegraphics[width=\textwidth]{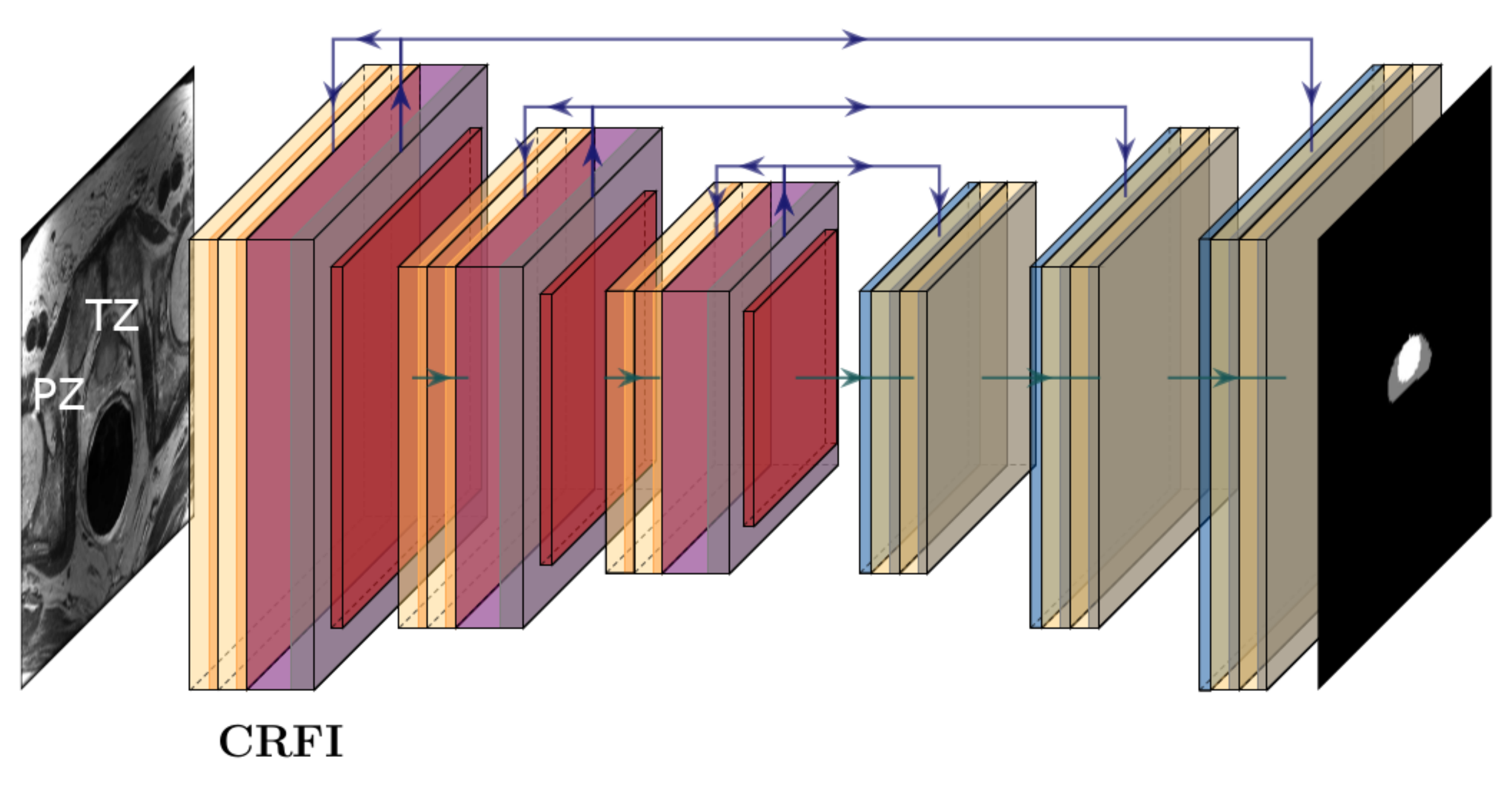}
\caption{}
\end{subfigure}
\\           
\begin{subfigure}[b]{0.9\textwidth}
\includegraphics[width=\textwidth]{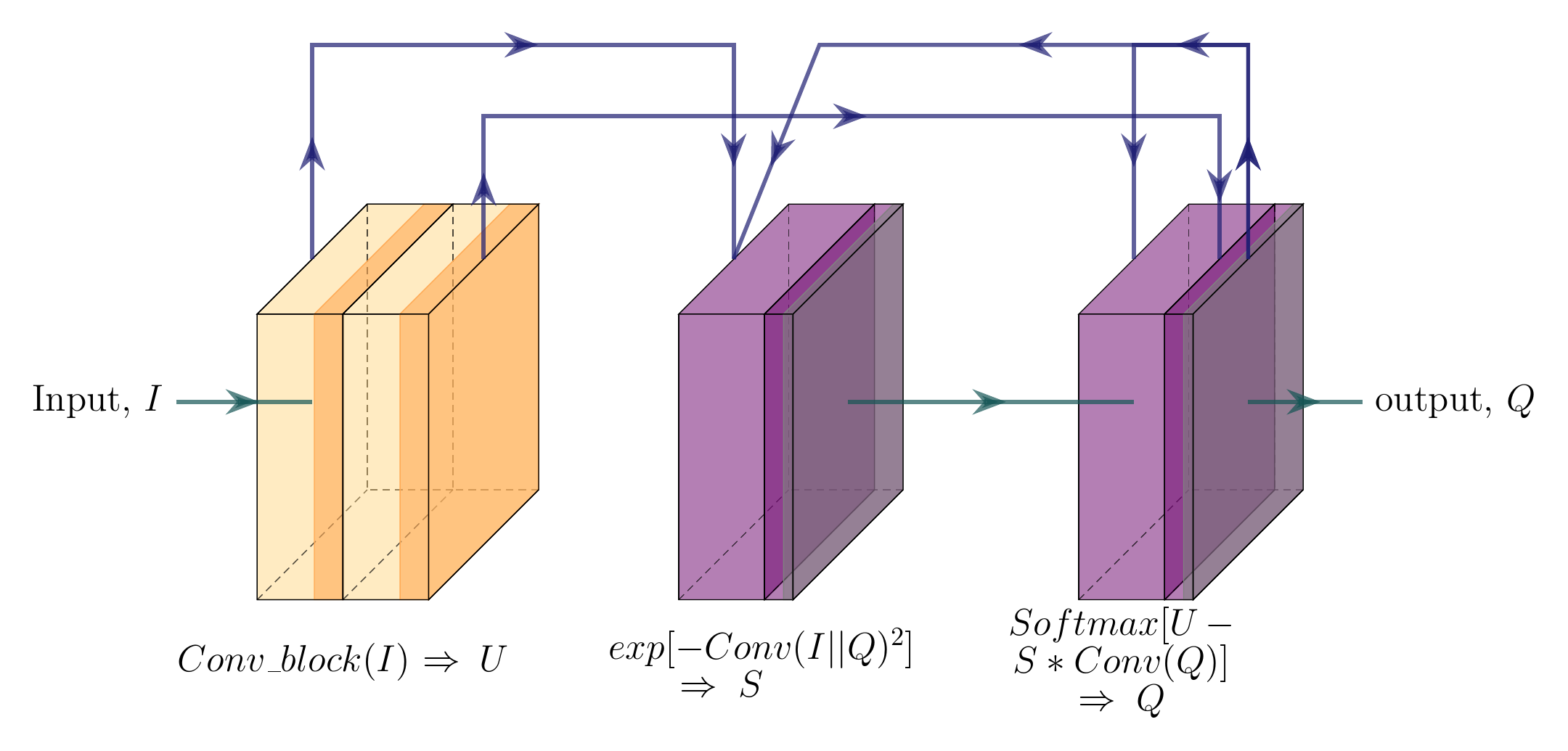}
\caption{}
\end{subfigure}
\end{tabular}
\caption{(a) Modified CRF inserts (CRFIs) in SegNet (SegNet+CRFIs), applied to the encoder side of SegNet. Colors are yellow for convolution block, magenta for CRFI, red for 2$\times$2 max-pooling, and blue for upsampling and concatenation. Apart from SegNet$'$s bypath connections, CRFI has feedback connections that encourage the data consistency or similarity at multiple levels. (b) Diagram of the connection between a convolution block (yellow) and the recurrent CRFI (magenta). The CRFI combines the input, $I$, and convolution block output $U$ based on the spatial similarity in $I$. Intermediate variables are $S$ for similarity weighting of the input image and $Q$ for conditional probability from the previous iteration. $S$ is computed by a convolution layer applied to the input $I$ with a new activation function, i.e., $S=exp[-Conv(I||Q)^2]$. Small $S$ corresponds to high locally similarity, which results in a small penalty on the pairwise potential in Eq. \eqref{eq:1}. $Q$ for the conditional probability that would produce the probable output during iteration for similar intensity and adjacent pixels to be labeled with the same output vector. The following step, i.e., $S*Conv(Q)$, generated the pairwise part, i.e., $Softmax[U-S*Conv(Q)]$, updating $Q$. The darker band in some boxes indicate the activation function.}
\label{fig:fig1}

\end{figure}

In this study we set M=1, i.e., one Gaussian kernel, and we reduced the message passing in Algorithm \ref{alg:1} to an element-wise multiplication of $S$ and the local sum of $Q$. $S(x_i)$ was the approximation of $S_{m}(x_i,x_j)$ for adjacent pixels within a convolution kernel, i.e., $S(x_i)$ was derived from the average local distance map $d$, resulted that $\sum_{j\neq i} S_{m}(x_i,x_j) Q_i(l)$ was replaced by $ S(x_i)\sum_{j\neq i, j \in \Omega(i)} Q_i(l)$ for a truncated local summation for  $j \in \Omega(i)$ in message passing. Then a convolution layer can perform such summation operation locally for $Q_i(l) $, as well as for learning weightings in pre-weighting and the compatibility transform steps, i.e., these steps can be approximated by $Conv(Q)$. This simplification allowed the CRF to be implemented fully by two convolution layers, one for $P=S*Conv(Q)$, an element-wise multiplication of $S$ and $Conv(Q)$, and another for $d=Conv(Q)$, plus a softmax and newly introduced activation function, i.e., $S=exp(-d^2)$ from Eq. \eqref{eq:2}, as shown in Figure \ref{fig:fig1}b. This simplified CRF can be inserted in the UNet or SegNet structure without too much increase in the computation burden, as shown in Figure \ref{fig:fig1}. The simplified CRFI algorithm is summarized below:
\\

\noindent
{Algorithm 2, Simplified neural network implementation for Algorithm Algorithm 1, which can be approximately by two convolution layers.} \\
$ U = Conv\_block(I)$ \null\hfill Convolution block in UNet/SegNet \\
$ Q = Softmax(U)$ \null\hfill Initialization \\
$ {While (not\_converged}$ \\ 
$  d=Conv(I)$ \null\hfill Average local distance measurement\\
$  S=exp(-d^2)$ \null\hfill  Similarity weighting \\
$  P = S*Conv(Q)$ \null\hfill Message passing, pre-weighting, and compatibility transform \\
$  \tilde{Q} = U-P $                       \null\hfill Add unary\\
$  Q =Softmax(\tilde{Q}) $                    \null\hfill Softmax  \\
\label{alg:2}

\begin{figure}[h!]
\centering

\includegraphics[width=\textwidth]{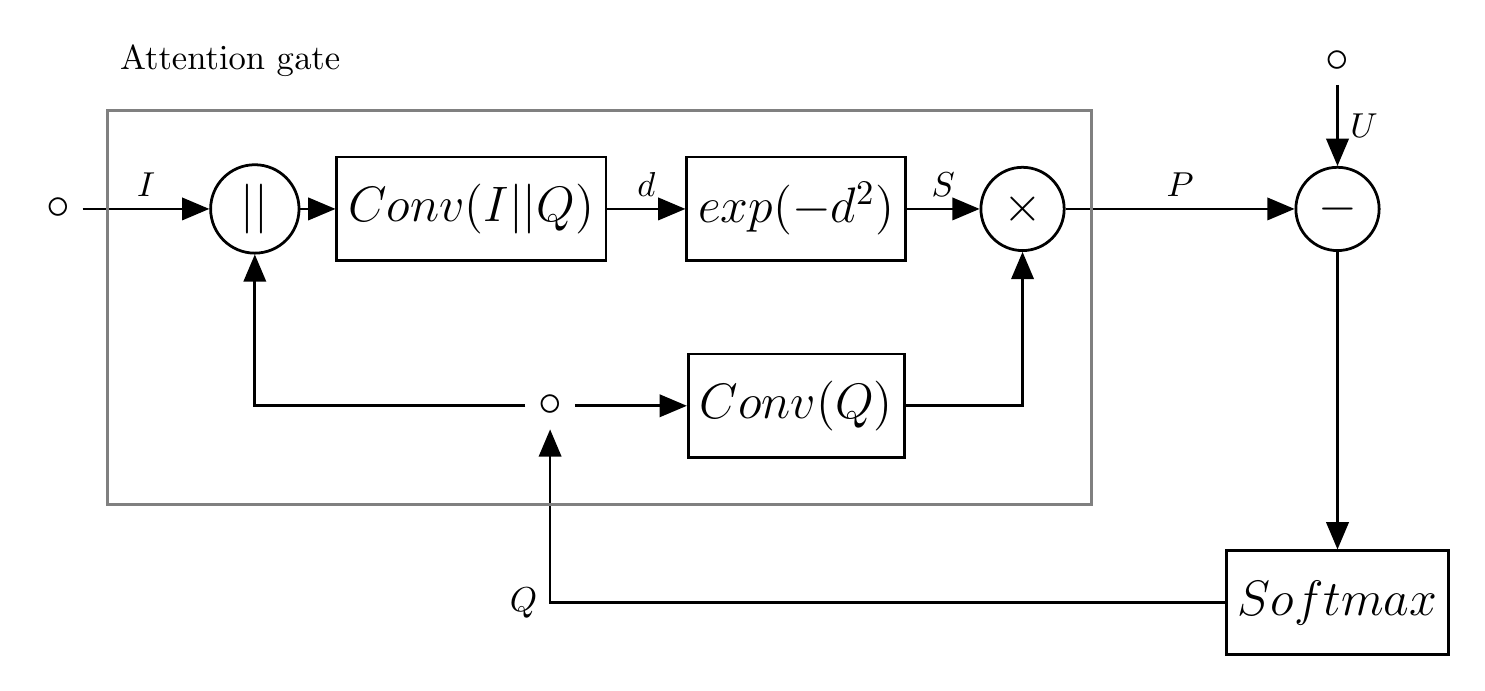}
\caption{ Data flow of the recurrent CRFI in Figure \ref{fig:fig1}. The CRFI combines the input, $I$, and convolution block output $U$ based on the spatial similarity measured on $I$.  $U$ is the negative of the unary energy in CRF. $S=exp(-d^2)$ is similarity weighting from the average local distance map $d=Conv(I||Q)$, and $Q$ for conditional probability from previous iteration. The whole CRFI can be viewed as an attention gate inside a CRF. The $S$ is gated by $Conv(Q)$ through element-wise multiplication, i.e., $P = S*Conv(Q)$. }
\label{fig:fig2}
\end{figure}

Another practical consideration was to utilize the attention gate for mean-field iteration \cite{Oktay2018a}, in order to insert the CRF in a UNet or SegNet structure without creating hurdles in training the network end-to-end. We considered the simplified message passing step, i.e., $S*Conv(Q)$ (where $*$ for element-wise multiplication) in Algorithm \ref{alg:2}, was an attention gate for $S$, gated by $Conv(Q)$, in Figure \ref{fig:fig2}. This was similar to the attention gate for RNN in \cite{Oktay2018a}, as a series of multiplications of the attention maps and the input feature maps. In \cite{Oktay2018a}, a special design that concatenated the gating $g$ and input feature map $x$, i.e., $g||x$, was used to gate $x$, instead of directly using $g$, which may help the gradient backpropagation in this RNN. To adapt this idea, we simply replaced the average local distance measurement in Algorithm \ref{alg:2} by
\begin{equation}
	d=Conv(I||Q)
	\label{eq:6}
\end{equation}
The complete data flow of the modified CRF is illustrated in Figures \ref{fig:fig1} and \ref{fig:fig2}.


To reiterate the method above, as illustrated in Figure \ref{fig:fig1}, in this study, three simplified conditional random field layers as inserts (CRFIs) were added to the original SegNet structure. The CRFI combines the input and output of the convolution block, i.e., $I$ and $U$, based on the spatial local similarity measured on $I$. As explained in Algorithm \ref{alg:2}, we replaced the message-passing step by a convolution layer so that the CRFasRNN structure can be inserted into each level of SegNet on the encoder side (Fig. \ref{fig:fig1}b). In this CRFI structure, as shown in Figure \ref{fig:fig1}b, intermediate variables were: $U$ for negative of unary energy, $S$ for Gaussian similarity weighting from the average local distance map, $d$, and $Q$ for probability from the previous iteration.  These CRFIs on encoder side of the SegNet convert input feature vectors from convolution blocks, $U=Conv\_block(I)$, into consistent output feature vectors based on the similarity measured on $I$, as governed by the exponential activation, $exp(-d^2)$, and a convolution layer (i.e., Eq. \eqref{eq:6}) that determined the average local distance map, $d$. For example, small $d$ correspond to high local similarity, which can result in a large $exp(-d^2)$ and a small penalty on the pairwise energy defined in Eqs. \eqref{eq:1} and \eqref{eq:4}. The CRFasRNN method then used the Gaussian kernel, i.e., $exp(-d^2)$ as the weighting for the ``bilateral filter'' for $Q$, in Algorithm \ref{alg:2} to follow the mean-field inference method with the pairwise energy defined in Eq. \eqref{eq:4}. With two simplifications, i.e., $M=1$ and Gaussian weighting was applied locally and approximately, a convolution layer could be used to replace several steps in Algorithm \ref{alg:1}. This setting allowed the use of two convolution layers to perform the simplified CRF. Such CRFI can be placed either on the encoder side of SegNet or at the end of the SegNet, as to how CRFasRNN was used \cite{Zheng2015a} and can be compatible with the end-to-end training of SegNet. In addition, vector similarity in Eq. \eqref{eq:2} was defined as the existence of small average distance in spatial adjacent feature vectors, which was implemented using a convolution layer in this study.\par

\subsection*{Training and testing}
As an initial experiment, 725 2D images (i.e., 29 MRI cases) were used in training; while, 174 2D images (i.e., 6 cases) were used in testing. The data augmentation methods used included $\pm$ 30-pixel shift, $\pm$ 5-degree 3D rotation, $\pm$ 5\% affine scaling, $\pm$ 50\% intensity scaling, and $\pm$ 10\% additive noise. For the training of all networks, the loss function (LF) was defined as a joint negative logarithm of the soft Dice coefficient: (2*intersection)/(count(label)+count(prediction), and weighted cross-entropy loss. For weighted cross-entropy loss, 0.02, 1.0, and 1.0 weightings for background, TZ, and PZ classes were used, respectively. ADAM operator was used with a fixed training rate of $10^{-4}$. Within the CRFI, $5\times 5$ convolution kernels were used for two convolution layers, as shown in Figure \ref{fig:fig1}b. The number of iterations for CRFI was set to 5 in Algorithm 2 in this study.
All the neural networks were implemented in Tensorflow software v1.0 (https://www.tensorflow.org/).\par

\section*{Results}
\begin{figure}[h!]
\includegraphics[width=\textwidth]{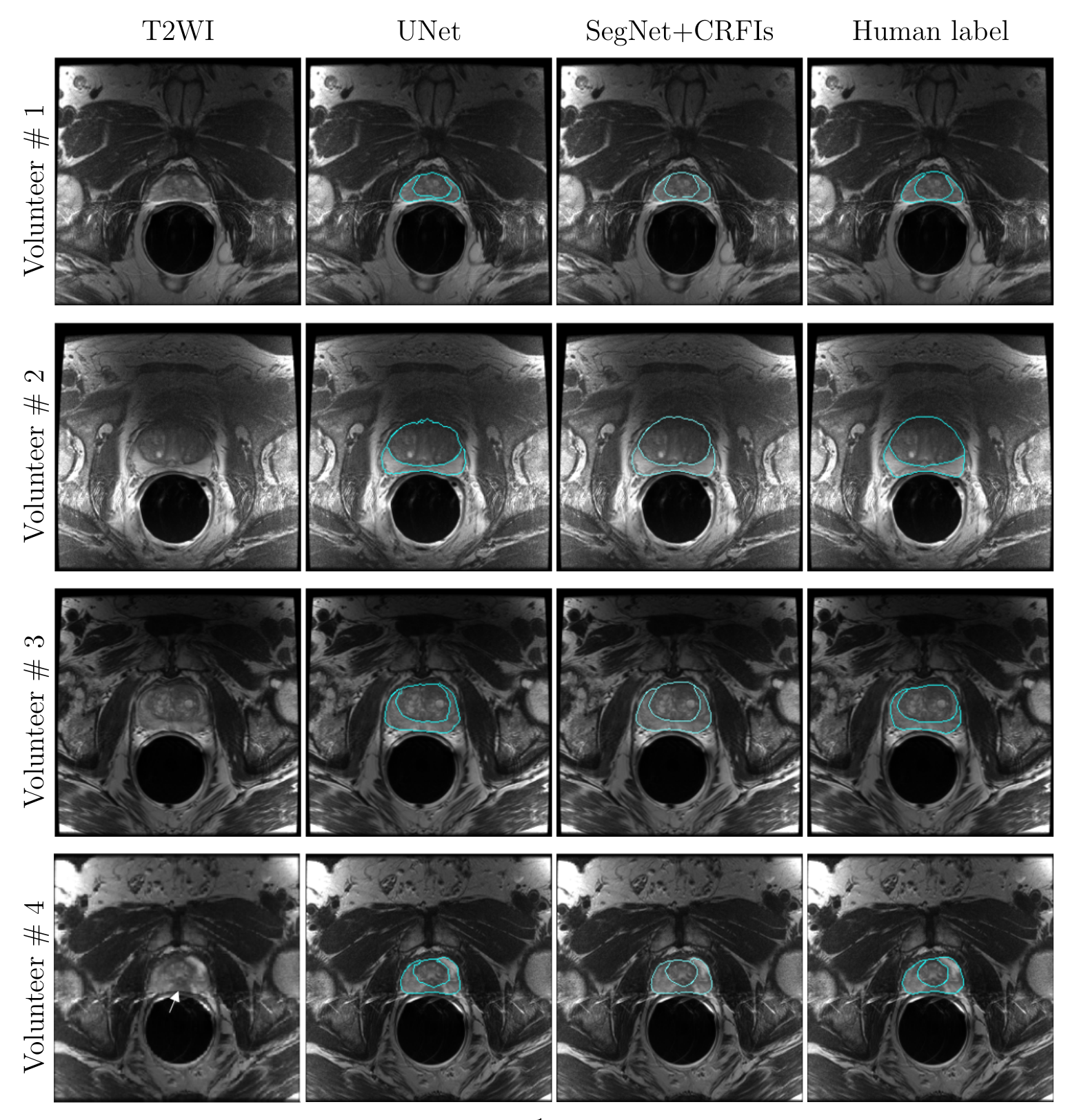}
\caption{Representative neural network segmentation results from 4 participants. The neural network predictions and the human-labeled PZ and TZ were contoured and overlaid on the T2 weighted MR images. Note that the neural network was able to predict the correct zonal boundary in most cases, even at the presence of T2 lesion (arrow). }
\label{fig:fig3}
\end{figure}

\begin{figure}[h!]
\centering
\includegraphics[width=\textwidth]{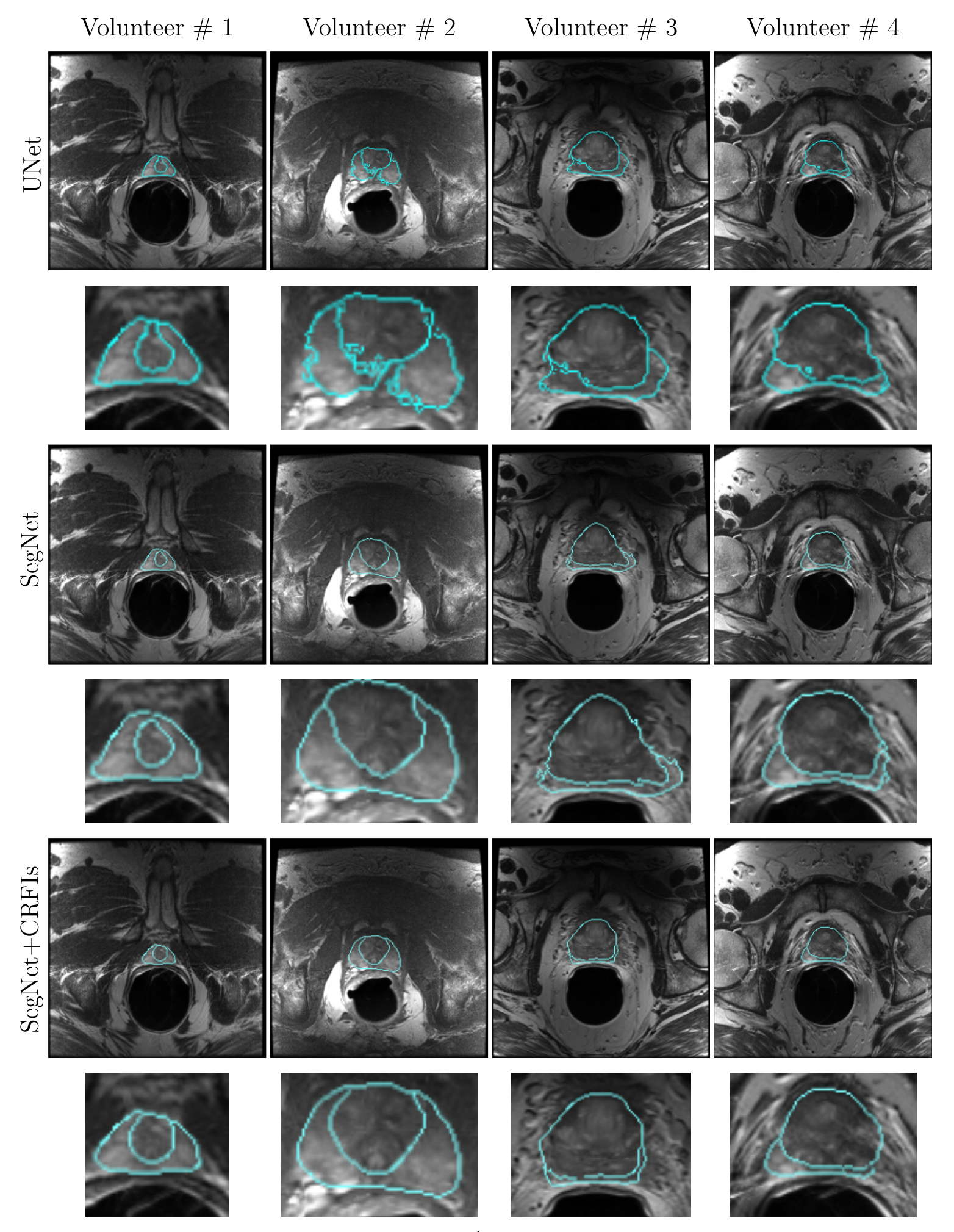}
\caption{Comparison of UNet, SegNet, and proposed SegNet+CRFIs segmentation and zoom-in views overlaid on the T2-weighted images. Slices were from the top or bottom of the prostate, where the segmentation was challenging. Results indicate the ill-defined boundaries in UNet and SegNet outputs. The SegNet+CRFIs segmentation has higher Dice score compared with SegNet (Table 1) with smooth boundaries of the prostate.}
\label{fig:fig4}
\end{figure}

Figure \ref{fig:fig3} shows typical neural network segmentation results from 4 participants. The neural network was able to predict the correct zonal boundary in most cases. The SegNet+CRFIs can also provide slightly better contour detection/interpolation compared with that of UNet (e.g., in Volunteer \#2 in Fig. \ref{fig:fig3} anterior surface of TZ). SegNet+CRFIs also had slightly better Dice score compared with that of UNet (Table \ref{tab:1}). 
Figure \ref{fig:fig4} compared the UNet, the SegNet alone, and proposed SegNet+CRFIs segmentation with a focus on the challenging slices for UNet on 4 participants. The results in Figure \ref{fig:fig4} showed the ill-defined boundaries in UNet and SegNet outputs. The SegNet+CRFIs segmentation showed the robustness in determining smooth boundaries of the prostate compared with the baseline UNet and SegNet segmentation.\par
\begin{figure}[h!]
\centering
\includegraphics[width=\textwidth]{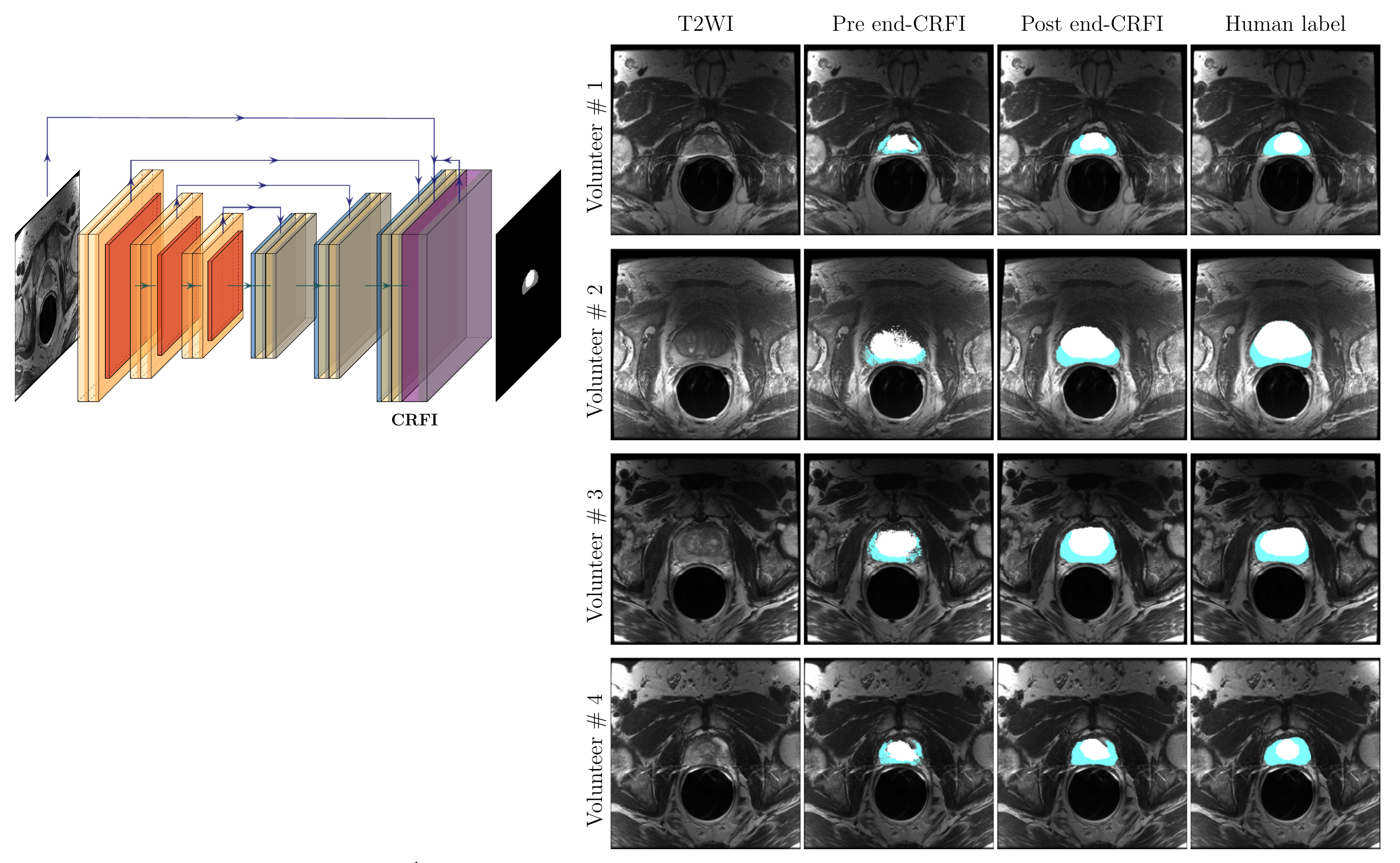}
\caption{(Top) Scheme of a SegNet with a CRFI at the end. (Bottom) Representative neural network segmentation results from 4 participants. Predictions before or after the CRFI. Note that the CRFI was able to correct the errors from the SegNet, i.e., post vs. pre end-CRFI, resulting in smooth and consistent segmentation for prostate.}
\label{fig:fig5}
\end{figure}
\begin{figure}[h!]
\centering
\includegraphics[width=\textwidth]{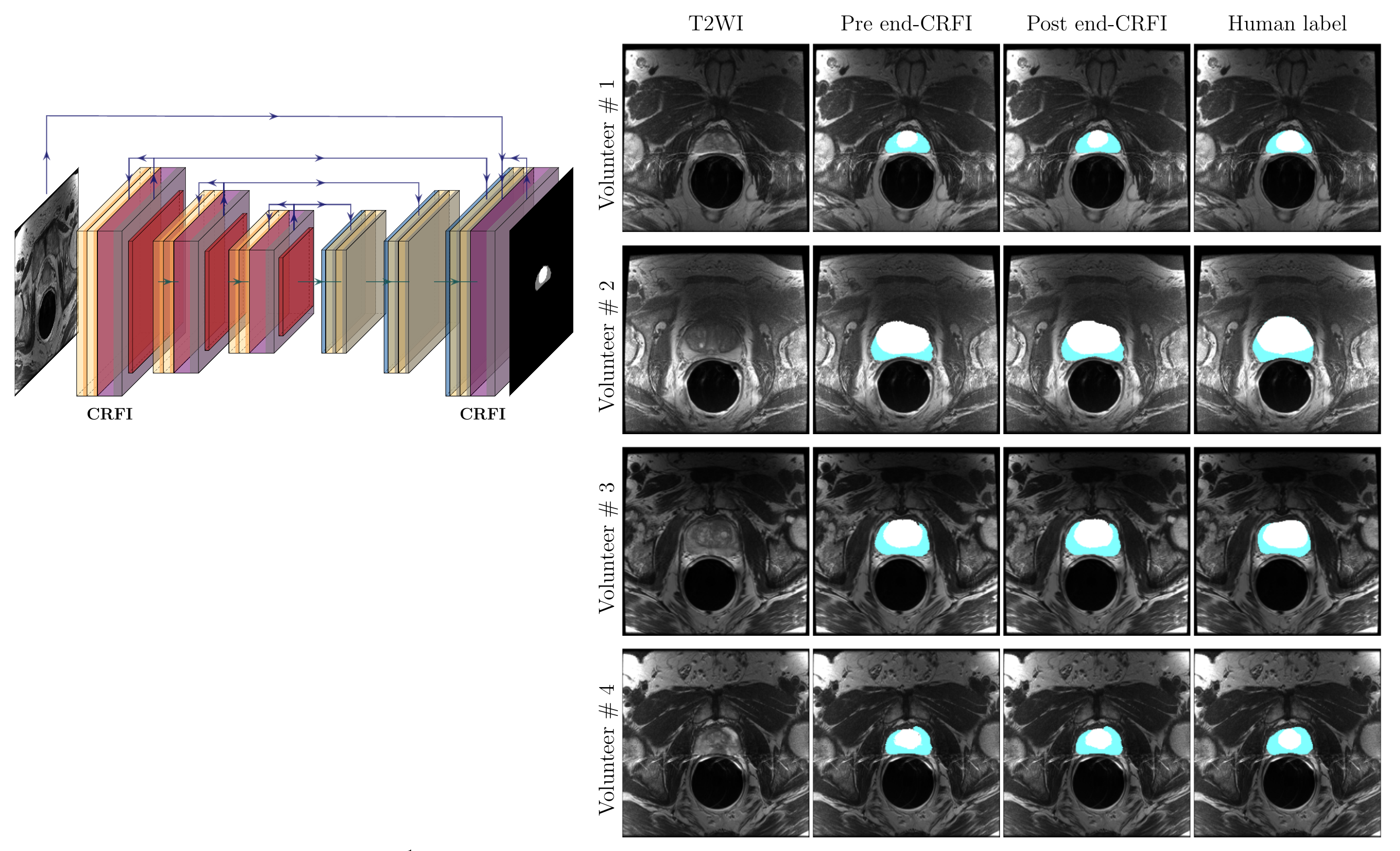}
\caption{(Top) Scheme of a combination of two methods: CRFIs on encoder side and at the end. (Bottom) Predictions before or after end-CRFI and the human-labeled PZ and TZ were overlaid on the T2 weighted MR images. Note that the segmentation from pre and post end-CRFI were both smooth and consistent. Meanwhile, results from post end-CRFI had highest Dice score in this study.}
\label{fig:fig6}
\end{figure}

Figure \ref{fig:fig5} shows another experiment using a CRFI at the end of SegNet. The CRFI was able to correct the errors from the SegNet output, i.e., post vs. pre end-CRFI, resulting in smooth and consistent segmentation for prostate. This result confirmed that the functionality of CRFI was to encourage the consistency of the output labels with reference to the input image. Figure \ref{fig:fig6} shows a similar experimental setting, but with CRFIs on the encoder side of SegNet and at the end, the comparison was performed for results from before and after the end-CRFI. Interesting, the results were largely the same, suggesting the CRFIs on encoder side of SegNet, i.e., SegNet+CRFIs, was effective in preserving the smooth contours and it was compatible with having a CRFI at the end of the network and an end-to-end training strategy.

\section*{Discussion}
These CRF inserts may help the convergence of the SegNet during training because of the multi-scale local spatial similarity is encouraged, i.e., multi-level attention focusing that is similar to methods in (15,16). The functionality of CRFI at the end of SegNet was similar to the conventional CRFasRNN  \cite{Zheng2015a} or DenseCRF \cite{krhenb2012}. Meanwhile, in contrast with the previous approaches, proposed CRFI can be placed inside the UNet or SegNet structure as the SegNet+CRFIs combination demonstrated in this study. The SegNet+CRFIs also achieved a higher Dice coefficient compared with baseline UNet or SegNet. Intuitively, for each CNN block plus CRFI, the combination behavior was like that of CNN + CRF, i.e., preserving the edge of the object. The CRFI in this setting encouraged the multi-level spatial similarity on the encoder side and helped the edge detection/interpolation and model convergence. 
The proposed SegNet + CRFI deep neural network can perform zonal prostate segmentation, achieving higher Dice coefficients compared to those provided by non-neural network state-of-the-art methods, i.e., Dice coefficients of 0.60 and 0.70 for PZ and CG/TZ by template-based and C-means clustering \cite{Chilali2016} and level-set \cite{Toth2014}. The limitations of the current study include the small sample size and the homogeneous MRI exam, i.e., all using the same parameters and from one institute and all patients bear untreated prostate cancer. Further testing our method on a larger set of clinical and multi-center data is required to confirm these results. \par
This study also presented a feasible simplification for CRF and mean-field inference, as we used two convolution layers to implement the CRFI in the proposed model, allowing the CRFI to be placed either on the encoder side of SegNet or at the end of it or both. 
Interestingly, the similarity or the average local distance map in proposed CRFI might have the representation capacity for affine transform on adjacent feature vectors, which was used globally in CapsuleNet \cite{Sabour2017}. 
We also empirically found that the SegNet+CRFIs network could recognize the relative spatial relation between TZ and PZ at the early stage of the training, leading to a rapid training convergence compared with the baseline Unet. 
\par
\section*{Conclusion}
In summary, three fully convolutional neural networks based on SegNet and CRF for zonal prostate segmentation were presented and demonstrated high accuracy. The SegNet with CRFI solved the ill-defined boundary with UNet segmentation and achieved the higher Dice coefficient compared with baseline UNet and SegNet. This study demonstrated that adding CRFI to the SegNet was appropriate augmentation for improving the segmentation of the prostate. \par

\section*{Acknowledgement}
This research was supported by funds from the California Tobacco-Related Disease Research Grants Program Office of the University of California, Grant Number: 28IR-0060 and by funds from the NIH: R01 CA148708, R01 EB16741, and from TRDRP 131866A and American Cancer Society Research Scholar Grant RSG-18-005-01 CCE.
\vspace{1cm}
\begin{table}[H]
	\caption{Dice coefficients measured on six T2W MRI cases (N = 71, i.e., 71 2D images on the prostate, mean $\pm$ standard deviation).}
	\centering
	\begin{tabular}{c c c c}
		\hline \hline
	        & Peripheral zone &	Central gland & Whole gland\\ 
		\hline 
		UNet & 0.734 $\pm$ 0.149 & 0.825 $\pm$ 0.127 & 0.851 $\pm$ 0.034\\ 
		
		SegNet & 0.736 $\pm$ 0.176 & 0.822 $\pm$ 0.144 & 0.867 $\pm$ 0.030\\ 
		SegNet+CRFIs & 0.737 $\pm$ 0.167 & 0.829 $\pm$ 0.131 & 0.888 $\pm$ 0.029 \\ 
		SegNet, post end-CRFI & 0.744 $\pm$ 0.148 & 0.847 $\pm$ 0.067 & 0.852 $\pm$ 0.058 \\
		SegNet+CRFIs, post end-CRFI & 0.757 $\pm$ 0.135 & 0.842 $\pm$ 0.091 & 0.890 $\pm$ 0.022 \\
		\hline 
	\end{tabular}
	\label{tab:1}
\end{table}

\clearpage

\clearpage
\bibliographystyle{plain}
\bibliography{references}
\end{document}